\documentclass{article}

\usepackage{arxiv}

\usepackage[toc,page]{appendix}

\usepackage{graphicx}
\usepackage{dcolumn}
\usepackage{bm}

\usepackage[mathlines]{lineno}

\usepackage[utf8]{inputenc}
\usepackage{gensymb} 
\usepackage{booktabs} 
\usepackage{subcaption} 
\usepackage{color} 
\usepackage{adjustbox}
\usepackage{amsmath}
\usepackage{mathtools, nccmath}
\usepackage{xparse}
\DeclarePairedDelimiterX{\set}[1]{\{}{\}}{\setargs{#1}}
\usepackage{listings}
\usepackage{enumitem}
\definecolor{codegreen}{rgb}{0,0.6,0}
\definecolor{codegray}{rgb}{0.5,0.5,0.5}
\definecolor{codepurple}{rgb}{0.58,0,0.82}
\definecolor{backcolour}{rgb}{0.95,0.95,0.92}

\lstdefinestyle{mystyle}{
  backgroundcolor=\color{backcolour}, commentstyle=\color{codegreen},
  keywordstyle=\color{magenta},
  numberstyle=\tiny\color{codegray},
  stringstyle=\color{codepurple},
  basicstyle=\ttfamily\footnotesize,
  breakatwhitespace=false,         
  breaklines=true,                 
  captionpos=b,                    
  keepspaces=true,                 
  numbers=left,                    
  numbersep=5pt,                  
  showspaces=false,                
  showstringspaces=false,
  showtabs=false,                  
  tabsize=2
}

\usepackage[utf8]{inputenc}
\usepackage[english]{babel}
\usepackage{graphicx}
\usepackage{gensymb} 
\usepackage{booktabs} 

\usepackage{lipsum}

\usepackage[normalem]{ulem} 
\usepackage{bm}
\usepackage{bbm}
\usepackage{amssymb}
\usepackage{multirow}

\usepackage{url}
\usepackage[hidelinks]{hyperref}

\usepackage{algcompatible}
\usepackage{newfloat}
\DeclareFloatingEnvironment[
    fileext=loa,
    listname=List of Algorithms,
    name=ALGORITHM,
    placement=tbhp,
]{algorithm}

\title{OpenTPS - Open-source treatment planning system for research
in proton therapy}

\usepackage{authblk}

\author[1]{Sophie Wuyckens}
\author[2]{Damien Dasnoy}
\author[2,3]{Guillaume Janssens}
\author[2,3]{Valentin Hamaide}
\author[1]{Margerie Huet}
\author[2]{Estelle Loÿen}
\author[2]{Gauthier Rotsart de Hertaing}
\author[2]{Benoit Macq}
\author[1,4,5]{Edmond Sterpin}
\author[1]{John Lee}
\author[1,3]{Kevin Souris}
\author[1]{Sylvain Deffet}

\affil[1]{Université catholique de Louvain, Institut de Recherche Expérimentale et Clinique (IREC), Molecular Imaging, Radiotherapy and Oncology, 1200 Woluwe-Saint-Lambert, Belgium}
\affil[2]{Université catholique de Louvain, Institute of Information and Communication Technologies (ICTEAM),  Louvain-La-Neuve, 1348, Belgium}
\affil[3]{Ion Beam Applications SA, Louvain-La-Neuve 1348, Belgium}
\affil[4]{KU Leuven, Department of Oncology, Laboratory of experimental radiotherapy, Leuven, 3000, Belgium}
\affil[5]{Particle Therapy Interuniversity Center Leuven - PARTICLE, Leuven, 3000, Belgium}

\renewcommand{\headeright}{Technical Report}
\renewcommand{\undertitle}{Technical Report}
\renewcommand{\shorttitle}{OpenTPS}

\begin{document}
\maketitle

\begin{abstract}
\textit{Introduction.} Treatment planning systems (TPS) are an essential component for simulating and optimizing a radiation therapy treatment before administering it to the patient. It ensures that the tumor is well covered and the dose to the healthy tissues is minimized. However, the TPS provided by commercial companies often come with a large panel of tools, each implemented in the form of a black-box making it difficult for researchers to use them for implementing and testing new ideas. To address this issue, we have developed an open-source TPS. \\
\textit{Approach.} We have developed an open-source software platform, OpenTPS (\url{opentps.org}), to generate treatment plans for external beam radiation therapy, and in particular for proton therapy. It is designed to be a flexible and user-friendly platform (coded with the freely usable Python language) that can be used by medical physicists, radiation oncologists, and other members of the radiation therapy community to create customized treatment plans for educational and research purposes. \\
\textit{Result.} OpenTPS includes a range of tools and features that can be used to analyze patient anatomy, simulate the delivery of the radiation beam, and optimize the treatment plan to achieve the desired dose distribution. It can be used to create treatment plans for a variety of cancer types and was designed to be extended to other treatment modalities. \\
\textit{Significance.} A new open-source treatment planning system has been built for research in proton therapy. Its flexibility allows an easy integration of new techniques and customization of treatment plans. 
It is freely available for use and is regularly updated and supported by a community of users and developers who contribute to the ongoing development and improvement of the software. 
\end{abstract}


\section{Introduction}
Thanks to the ballistic properties of heavy charged particles, proton therapy has the potential to better confine the dose to the target volume and spare healthy tissues more than conventional radiation therapy. In practice, however, the full potential of proton therapy is limited by its maturity gap with conventional radiation modalities. For example, the first 3D intensity modulated proton therapy (IMPT) treatment was delivered in 1999, at a time when volumetric arc therapy (VMAT) treatments were becoming popular in photon therapy \cite{Hatano2019}.  Similarly, while VMAT was first used on a patient in 1994 \cite{Mackie1993}, the first prototype of intensity modulated proton arc therapy was only tested in 2019 \cite{Li2019}. Despite challenges, proton therapy is widely acknowledged as a beneficial option for some patients undergoing radiation treatment. However, the full potential of proton therapy systems cannot yet be fully realized due to limitations in the software, particularly in the treatment planning aspect.

Over the past three decades, proton therapy has benefited from a growing body of research aimed at realizing its full potential. This has led to the development of many computational tools, such as \href{https://openreggui.org/}{REGGUI} \cite{Janssens2011}, \href{http://www.openmcsquare.org/}{MCsquare} \cite{Souris2016}, \href{https://e0404.github.io/matRad/}{matRad}, and MIROpt \cite{Barragan2018}, among others. These tools are widely available and have helped accelerate research in proton therapy. In the early 2020s, we embarked on a project to develop our own common research platform for the various radiation therapy projects in tha lab. Using the widely-available and user-friendly Python programming language, the platform was designed to serve as a flexible framework for both personal and collective research efforts. Additionally, the platform was developed to be compatible with the increasing significance of artificial intelligence and to facilitate the distribution of computational tasks across multiple computers.

We propose a treatment planning system (TPS) that is specifically designed for research, called openTPS. Its open-source code is publicly available under the Apache-2.0 license. The code is primarily written in python and designed to be easily extended and customized. The software is organized into two packages: the Core package, which is a library that defines data classes, data processing methods, and IO methods, and the GUI package, which offers a graphical interface for viewing and interacting with the data.

The Core package of OpenTPS is the main software library and includes a range of features that are essential for proton therapy treatment planning. Some of the key features available in the Core package of OpenTPS include:
\begin{enumerate}[label=\Alph*.]
    \item Data management and processing: importing, exporting and processing patient data.
    \item Dose computation: computing the dose from proton therapy treatment plans using fast Monte Carlo simulations with MCsquare.
    \item Treatment planning: creating treatment plans for individual patients, including the ability to define beam angles, energy levels, spot spacing, etc.
    \item Treatment evaluation: performing quality assurance checks on treatment plans, including the ability to evaluate robustness, compare doses, and assess the conformity of the treatment plan to the patient's anatomy.
    \item Data augmentation: creating artificial variations of patient data for robustness evaluation and for machine learning applications.
\end{enumerate}

In the next section, we detail these tools and give an example of their use for the development of new planning methods though an introduction to the 4D image support suite which is included in OpenTPS. Assurance tools for dose quality are discussed in section~\ref{sec:GUI}.

\section{GUI package\label{sec:GUI}}
\begin{figure*}
    \centering
    \includegraphics[width=\textwidth]{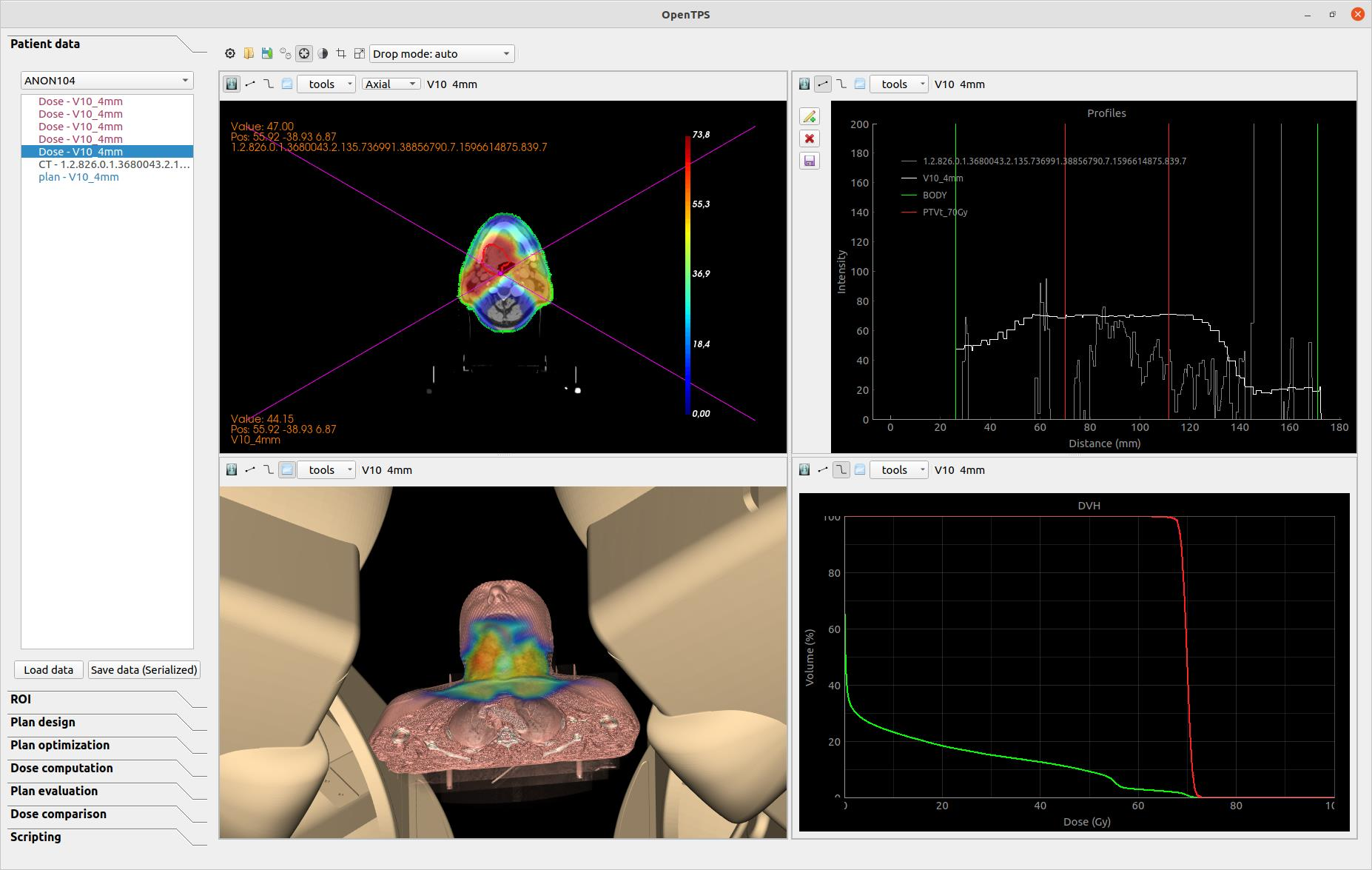}
    \caption{Main window of the OpenTPS GUI.}
    \label{fig:opentps_gui}
\end{figure*}

OpenTPS comes with an optional graphical user interface (GUI) for visualization of 3D images, dose maps, contours and dynamic images. Several other interactive tools are available: profile plots, visualization of DVH, segmentation, etc. The implementation is based on the PyQt, VTK, and PyQtGraph libraries, which are widely recognized for their capabilities in data visualization and analysis in scientific and medical applications. Figure \ref{fig:opentps_gui} shows a screenshot of the GUI loaded with a head-and-neck patient geometry and the optimized treatment dose associated to the case.
In addition, graphical modules have been implemented to design and optimize a treatment plan. The entire GUI is extensible and modifiable by the user. Examples of extensions can be found in the OpenTPS documentation.

\section{Core package}
\subsection{Data management and processing}
\label{dataManagAndProc}
The current version of OpenTPS supports the import patient data in DICOM format (CT, RT structures, RT dose, RT ion plan and vector field), MHD format (3D images), and OpenTPS-specific serialized object format. DICOM export is currently supported for treatment plans and doses. Any images can also be exported in MHD format.

The core library of OpenTPS includes a range of common image processing methods: resampling, coordinate system conversion, affine transformations, and filtering. Additionally, it has segmentation options such as automatic segmentation for body, lung and bones in CT images.

Rigid and deformable image registration methods are available, including the diffeomorphic Morphons algorithm \cite{Janssens2011}. This algorithm computes 3D to 3D deformations that are consistent with the anatomy of the organs. The deformations are represented by a velocity field and/or the more usual displacement field. While algebraic operations - such as summing and scaling fields - can be performed on velocity fields, they cannot be applied directly on displacement fields. On the other hand, when a deformation is applied on 3D data, the displacement field must be used. The displacement is automatically computed from the velocity field when the deformation is applied using a field exponentiation operation. 

\begin{figure*} 
\centering
\includegraphics[width=\textwidth]{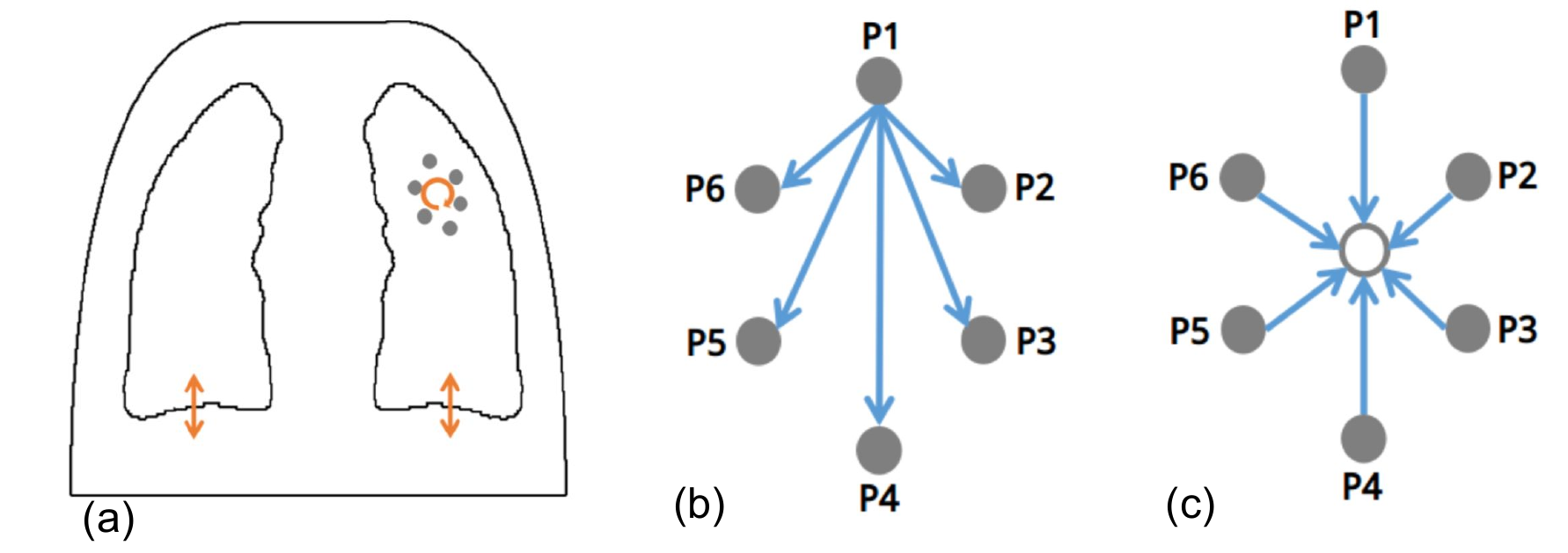}
\caption{The tumor motion is schematically represented by the hysteresis formed by the tumor position (gray circles) at each phase of the 4DCT (a). Deformation fields (blue vectors) are generated using registration between the first phase (P1) and all others phases (b). Then, all phases can be deformed to the average motion called the Mid- Position where the MidP-CT image is then created by taking the median in each voxel of all deformed phases (c).}
\label{midP}
\end{figure*}

The morphons registration is also used to create motion models. Deformation fields between the 4DCT phases are computed. Then with the deformations from all 4DCT phases, the midposition (MidP) image \cite{Wolthaus2008} can be computed, as illustrated in Figure \ref{midP}. A motion model, called Dynamic3DModels in OpenTPS, is composed of the MidP image and the various deformation fields from the MidP image to each of the 4DCT phases.

Digitaly reconstructed radiographs (DRR) can also be computed using a 3DCT to simulate in room x-ray devices.

\subsection{Dose computation}
OpenTPS was designed with the flexibility to support multiple dose engines as a way to provide users with a range of options for proton therapy treatment planning. It also allows users to take advantage of the strengths of different dose engines, such as the accuracy and precision of Monte Carlo-based engines, or the speed and efficiency of other types of engines.

\begin{figure*}
    \centering
    \includegraphics[width=\textwidth]{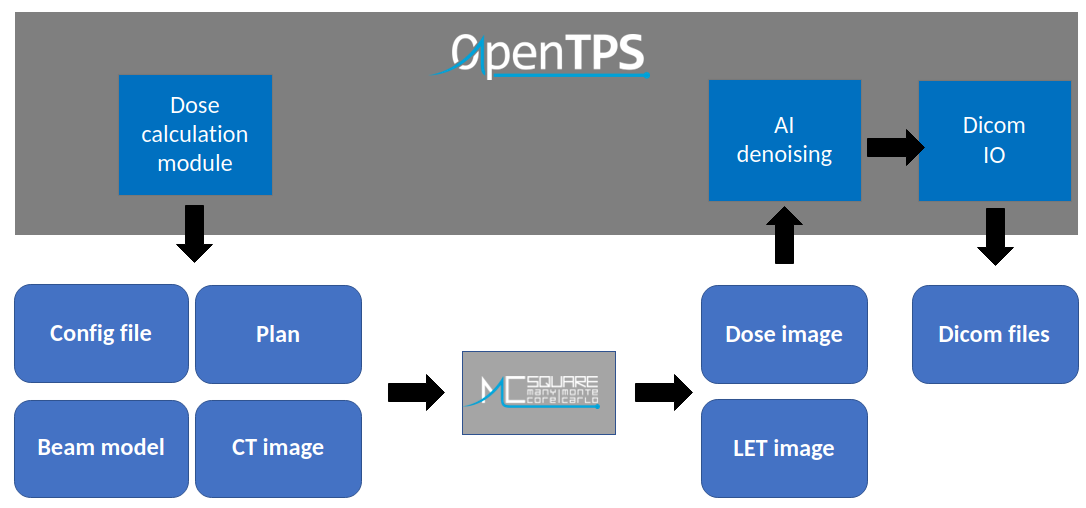}
    \caption{Example of dose workflow in OpenTPS}
    \label{fig:doseWF}
\end{figure*}

A stable release of the fast Monte Carlo named MCsquare is included in OpenTPS \cite{Souris2016}. It is available via the class named \textsf{MCsquareDoseCalculator} and can be used to compute dose distributions, dose influence matrices, and LET distributions. Input paramaters such as the number of primaries, the beam model and the CT calibration are required to compute a dose based on CT and a treatment plan. From those inputs, \textsf{MCsquareDoseCalculator} generates the files required by MCsquare. Then, MCsquare simulates the interaction of each particle in batches until it reaches a user-defined minimal statistical uncertainty or it simulates all the primaries. Mcsquare generates the output dose in mhd format in the OpenTPS workspace, a folder located by default in the home directory. 
Figure \ref{fig:doseWF} summarizes the dose calculation workflow in OpenTPS.

\subsection{Treatment planning}
The treatment planning module of OpenTPS provides a set of tools and functions that are used to create and optimize a treatment plan based on several inputs. In general, the method followed depends on the treatment delivery modality chosen. As OpenTPS is initially built for research in proton therapy, the implementation of the treatment planning module conform to IMPT delivery. In pencil beam scanning (PBS), the dose is painted using scanning magnets that steer each pencil beam to a specific location in the target volume, namely a \textit{spot}. Dividing the tumor into iso-energy layers, the treatment is therefore delivered by irradiating the volume spot-by-spot and layer-by-layer. This process is repeated for each irradiation direction. The intensity of each pencil beam being modulated, this treatment mode results in a very high dose conformity and a large flexibility in the planning allowing for better organ at risk sparing compared to others such as scattering techniques \cite{paganetti2018proton}. The modulation of the spot intensity is the aim of plan optimization.

\subsubsection{Plan optimization}
The treatment planning module operates in several steps. The first stage consists in creating a \textsf{PlanDesign} object. Loading a \textsf{CTImage} and \textsf{RTStruct} associated to it, the planner defines the gantry and couch angles for the given disease site. Based on this information, spots are placed in beam-eye-view on a hexagonal grid through a ray-tracing procedure that computes min-max water-equivalent path lengths (translated into energies) covering the target volume and then spaced by pre-defined spot and layer spacings for each beam direction. Internally, a \textsf{RTPlan} object is initialized with the angles, energy layers and spot positions. The spot intensities are left behind for the moment and initialized to one. Before proceeding to the optimization, the dose distribution of each spot onto the patient voxelized-geometry, namely a \textit{beamlet}, must be computed. By default, the Monte Carlo dose engine named MCsquare is used to simulate, for each spot, the physics of proton interactions with the matter and score the dose deposited in each voxel. Nevertheless, the framework is flexible enough to implement any other dose calculation engine. The output of this step is a beamlet matrix, named $\mathcal{A}$, which is stored in memory. To conserve memory space and enhance computational efficiency during optimization, a sparse matrix format is utilized. The number of rows in $\mathcal{A}$ corresponds to the number of voxels in the patient (resampled or not) and the number of columns corresponds to the number of spots in the plan. Finally, the clinical objective are added to the \textsf{PlanDesign} object. The region of interest (ROI), the metric (min, max or mean), the dose limit value and the objective weight must be specified for each \textsf{FidObjective}. An optional parameter can be set to make the objective robust against uncertainties.

\begin{figure}
    \centering
    \includegraphics[width=\textwidth ]{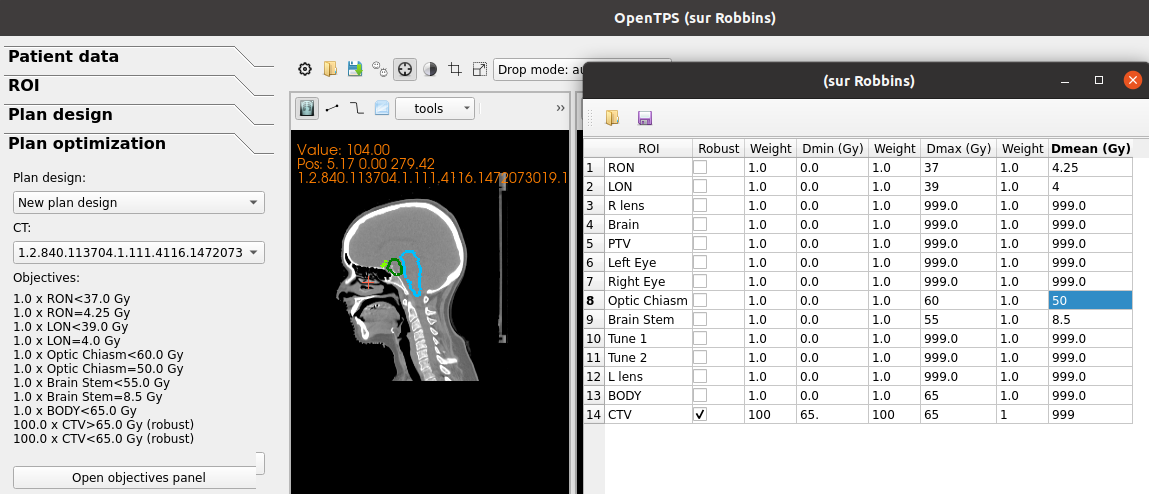}
    \caption{Objectives panel of the OpenTPS GUI}
    \label{fig:objectives}
\end{figure}

When the plan and the associated beamlets are created, the second stage of treatment planning can now take place. Knowing that the total dose delivered to the patient can be computed as the sum of all beamlets weighted by the intensity of the corresponding spots, the remaining task is to optimize the weights $\bm{x}$ so that the total dose $\bm{d}$ meets the optimization objectives. The optimization problem is stated as follows:
\begin{align}
\begin{split}
    \min_x f(\bm{d})\\
    \text{s.t.} \hspace{2em}
&   \bm{d} = \mathcal{A}\bm{x}\\
&   \bm{x} \geq \bm{0}
\end{split}
\end{align}

The total objective function $f(\bm{d})$ is defined as a weighted sum of asymmetric quadratic objectives with squared ramps used as surrogate to constraint on a minimum, maximum or on a mean dose. The total objective function has therefore usually four terms corresponding to the targets ($T$) and the OARs ($O$), which is written as follows:
\begin{multline}
    f(\bm{d}) = \sum_{t \in T} \left( \frac{1}{|v_t|} \sum_{i \in v_t} \left(w^+_t \cdot (d_i - p_t^{max} )_{+}^2 + w^-_t \cdot (d_i -p_t^{min} )_{-}^2\right) \right) \\+ \sum_{o \in O} \left( \frac{w^+_o}{|v_o|}  \sum_{i \in v_o} (d_i -p_o^{max} )_{+}^2 \right)  + \sum_{o \in O} \left( w_o  (\overline{\bm{d}} -p_o^{maxmean} )_{+}^2 \right)
\label{eq:objF}
\end{multline}
where $p$ is the prescribed dose and the notations $(\cdot)_+$ and $\overline{\bm{\cdot}}$ represent $max\{\cdot,0\}$ and $\frac{1}{|v|} \sum_{i \in v} d_i$ respectively. \\

OpenTPS comes with several built-in optimizers that solve the above problem. It is also designed so that other optimizers may be easily implemented. A variety of iterative solvers are provided: quasi-Newton's methods from Scipy, in-house implementation of LBFGS algorithms \cite{Byrd1995}, gradient-based methods such as the classical gradient descent and the FISTA algorithm \cite{beck2009fast} from PyUNLockBox \cite{pyunlocbox}. FISTA solves non-differentiable convex optimization problems by proximal splitting. For example, it can be of particular interest for ArcPT treatment plan optimization where one wants to maximize the spot sparsity by adding a $l_1$-norm penalty to the total objective function.

Linear programming \cite{dantzig2002linear} methods are also available and can be used to solve the optimization problem. Based on the list of objectives in the plan design, a model is built with linear equations instead of the quadratic version defined in Eq.~\ref{eq:objF}. The model and the solver are built on the Gurobipy package \cite{gurobi}, a Python interface to the powerful Gurobi solver, which requires a specific license. Gurobi includes a variety of algorithms to solve linear programming problems, including the simplex algorithm \cite{Dantzig1990}, the interior point method \cite{POTRA2000281}, and the concurrent optimizer. 

Continual research is being conducted to investigate new methods, such as the beamlet-free algorithm. This innovative approach simulates the dose in proton batches of randomly selected spots and evaluates their impact on the objective function in each iteration. Using an estimated gradient, the spot weights are then updated, leading to a new spot probability distribution from which the next spot will be sampled. This approach holds potential for further improving the treatment planning process in terms of time and memory requirements. 

The convergence rate of each of these algorithms may depend on the specific characteristics of the optimization problem. A summary of available solvers and their capabilities is provided in Table \ref{tab:algos}. The ``order" refers to the number of derivatives used in the optimization algorithm. First-order methods only use the first derivative (i.e., the gradient), while second-order methods use the first and second derivatives (i.e., the Hessian matrix).

\begin{table*}[t]
\begin{tabular}{|ll|l|l|ll|}
\hline
\multicolumn{2}{|l|}{\multirow{2}{*}{Solver}}                                                                & \multirow{2}{*}{Order} & \multirow{2}{*}{Solving function}                                                                                                   & \multicolumn{2}{l|}{Convergence}                                                                                                                                                     \\ \cline{5-6} 
\multicolumn{2}{|l|}{}                                                                                       &                        &                                                                                                                                     & \multicolumn{1}{l|}{Type}                         & Rate                                                                                                                             \\ \hline
\multicolumn{2}{|l|}{Gradient descent}                                                                       & 1                      & Smooth convex                                                                                                                       & \multicolumn{1}{l|}{Linear}                       & \begin{tabular}[c]{@{}l@{}}O(1/k)\\ where $k$: \# of iterations\end{tabular}                                                     \\ \hline
\multicolumn{2}{|l|}{FISTA}                                                                                  & 1                      & \begin{tabular}[c]{@{}l@{}}Smooth convex\\ Proximable non-smooth\end{tabular}                                                       & \multicolumn{1}{l|}{Linear}                       & \begin{tabular}[c]{@{}l@{}}O($1/k$) for smooth convex functions\\ O($1/k^2$) for proximable\\  non-smooth functions\end{tabular} \\ \hline
\multicolumn{2}{|l|}{\begin{tabular}[c]{@{}l@{}}Stochastic gradient descent\\ (``Beamlet-free")\end{tabular}} & 1                      & Smooth convex                                                                                                                       & \multicolumn{1}{l|}{/}                            & \begin{tabular}[c]{@{}l@{}}Depends on the problem, parameters\\ and learning rate.\end{tabular}                                  \\ \hline
\multicolumn{2}{|l|}{BFGS*}                                                                                  & \multirow{3}{*}{2}     & \multirow{2}{*}{Smooth convex}                                                                                                      & \multicolumn{1}{l|}{\multirow{3}{*}{Superlinear}} & \multirow{3}{*}{\begin{tabular}[c]{@{}l@{}}O($1/k^2$)\\ where k: \# of iterations\end{tabular}}                                  \\ \cline{1-2}
\multicolumn{2}{|l|}{L-BFGS*}                                                                                &                        &                                                                                                                                     & \multicolumn{1}{l|}{}                             &                                                                                                                                  \\ \cline{1-2} \cline{4-4}
\multicolumn{2}{|l|}{\begin{tabular}[c]{@{}l@{}}L-BFGS-B\\ (Scipy-only)\end{tabular}}                        &                        & \begin{tabular}[c]{@{}l@{}}Smooth convex\\ with bounds constraints\end{tabular}                                                     & \multicolumn{1}{l|}{}                             &                                                                                                                                  \\ \hline
\multicolumn{1}{|l|}{\multirow{3}{*}{LP**}}                         & Simplex                                & \multirow{3}{*}{/}     & \multirow{3}{*}{\begin{tabular}[c]{@{}l@{}}Linear objective functions \\ Linear equality and\\ inequality constraints\end{tabular}} & \multicolumn{1}{l|}{\multirow{3}{*}{/}}           & \begin{tabular}[c]{@{}l@{}}O($2^n$) - worst case\\ where n: \# of variables\end{tabular}                                         \\ \cline{2-2} \cline{6-6} 
\multicolumn{1}{|l|}{}                                              & Interior point                         &                        &                                                                                                                                     & \multicolumn{1}{l|}{}                             & Polynomial                                                                                                                       \\ \cline{2-2} \cline{6-6} 
\multicolumn{1}{|l|}{}                                              & Concurrent                             &                        &                                                                                                                                     & \multicolumn{1}{l|}{}                             & Depends on the problem                                                                                                           \\ \hline
\end{tabular}
\caption{Summary of available optimizers in OpenTPS. *Scipy and in-house versions available. **Gurobi optimizer.}
\label{tab:algos}
\end{table*}

\begin{figure*}
    \centering
    \includegraphics[width=\textwidth]{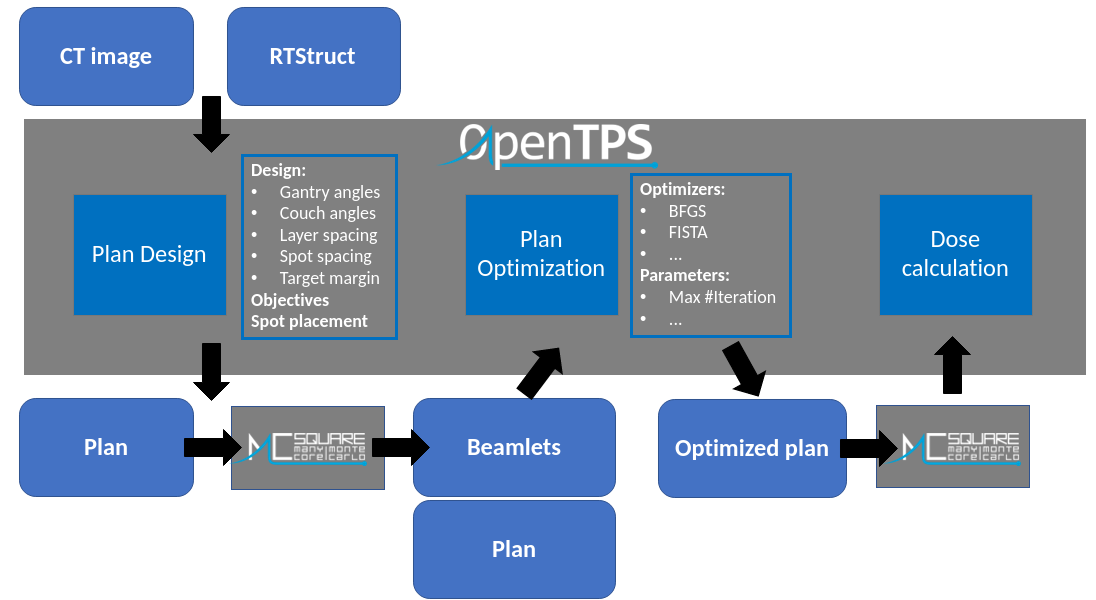}
    \caption{Optimization workflow in OpenTPS.}
    \label{fig:optiWF}
\end{figure*}

When the algorithm is selected and the plan is provided, an \textsf{IMPTPlanOptimizer} object is created. Additional parameters can be provided such as the desired stopping criterion for the optimization. The method \textsf{optimize} will start the optimization using the solver supplied. The result holds, among others, the optimized spot weights and the optimized dose computed from the beamlet matrix. Figure \ref{fig:optiWF} summarizes the different steps in the optimization workflow in OpenTPS.

The structure of the plan optimization module is divided in three main sub-modules. Firstly, the optimization engines described above are located in the \textsf{solvers} module. Secondly, to compose the objective function, several functions classes such as the classical dose objectives and more specialized objectives (norms, projections, etc.) are grouped in the \textsf{objectives} module. The user can also define a custom function for optimization by implementing an evaluation method and either a proximal operator or gradient method. Thirdly, acceleration schemes are included in the \textsf{acceleration} module such as backtracking linesearch typically used for gradient descent or Newton's methods, backtracking acceleration based on a quadratic approximation of the objective, FISTA acceleration for forward-backward solvers.

\begin{figure*}
    \centering
    \includegraphics[width=\textwidth]{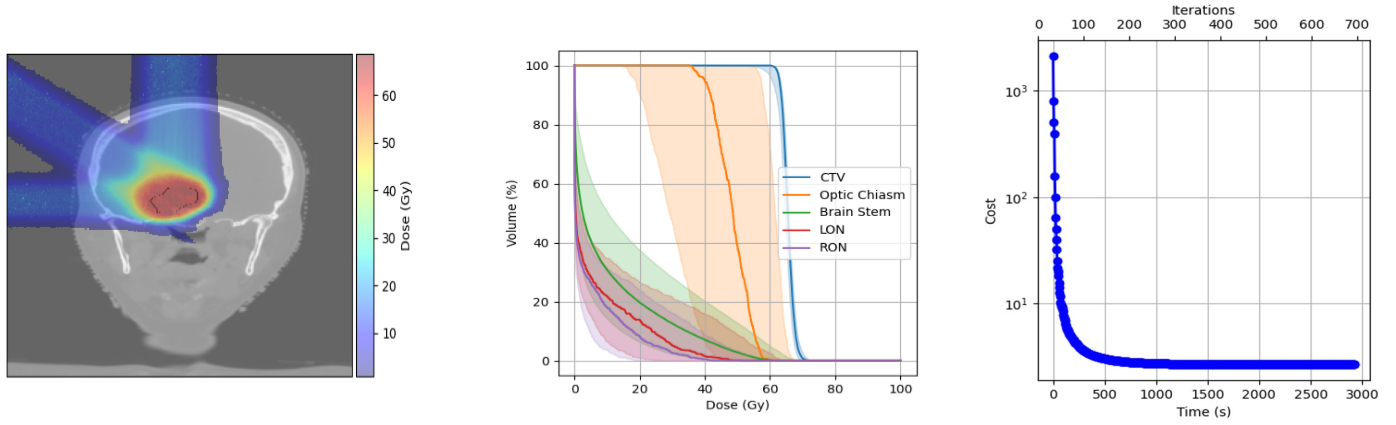}
    \caption{Robust optimized result obtained with scripting within the Core module. Left: dose distribution. Middle: DVHs. Right: Convergence plot}
    \label{fig:optiEx}
\end{figure*}

The GUI package of OpenTPS provides tools to analyze the resulting dose distribution. The dose can directly be overlaid using a color map on the CT image and contours to examine its homogeneity and conformity. Dose-volume histograms can also be generated to inspect quality of the plan. Dose line profiles can be obtained by manually drawing on the image in the viewer. Finally, DVH metrics or important index (conformity \cite{SHAW19931231} and homogeneity indices) can be extracted from the DVH. Figure \ref{fig:optiEx} shows an example of optimization result display for a base-of-skull tumor brain scripted in the core library. It includes a plot of the convergence of the objective function evaluation along the iterations.

\subsubsection{Robust optimization}
It is well-known that the PTV margin recipe \cite{VANHERK20001121} used in conventional photon treatments does not apply well in IMPT due to the presence of proton range uncertainties \cite{Lomax_2008,Unkelbach2009,Engwall2018}. The alternative approach is to make use of robust optimization \cite{Albertini_2011,Frederiksson2012}. In the robust scheme, the dose gradients per field are optimized so that they combine adequately even in the presence of range uncertainties. Specific combinations of uncertainties are commonly referred to as ``scenarios". They include patient motion and changes in anatomy resulting in variations in the range of the proton beam. The main idea behind robust optimization is the evaluation of the objective function in a discrete set of scenarios. Multiple robust optimization approaches have been proposed depending on the definition of the objective function and on the scenarios \cite{Unkelbach_2007,Unkelbach2009,Pflugfelder_2008,Fredriksson2011}. In OpenTPS, scenario-wise worst-case robust optimization is implemented. It minimizes the objective function in the worst scenario, within a certain confidence interval. The robust optimization problem, solved in OpenTPS, can now be summarized as,
\begin{align}
    \begin{split}
        \min_x \max_{s \in \mathcal{S}} f(\bm{d},s) \\
        \text{s.t.} \hspace{2em}
&   \bm{d} = \mathcal{A}\bm{x}\\
&   \bm{x} \geq \bm{0}
    \end{split}
\end{align}
where the max-operator applies to the scenarios $\{ s \in \mathcal{S}\}$.

The current implementation allows the selection of scenarios in the error space. The planner specifies the setup error (systematic and random) in millimeter as well as the range uncertainties in percentage. A total of 21 scenarios will be simulated (7 setup x 3 range). In terms of workflow, after the robustness settings are entered, the TPS will simulate the beamlets for each scenario. Next, the planner must specify among the objective list which ones are robust for the optimization. With that information, the optimizer will solve the problem by updating the weights using the objective function gradient calculated from the worst-case scenario at each iteration. The robust optimized treatment plan produces the best possible outcome according to the worst case scenario and robustness parameter settings. 

\subsection{Treatment evaluation}
Robust evaluation involves the examination of an optimized treatment plan under a variety of different scenarios that could occur during treatment delivery. This approach may be used to ensure that the treatment plan is effective and reliable under a range of possible conditions. 

\subsubsection{Robustness to setup and range uncertainties}
In OpenTPS, a dose-volume evaluation is performed to assess the quality of the treatment plan. Two strategies are available depending on the scenarios definition. The first method, which is based on good practice rules and is commonly used in clinical routine, involves verifying the dose in the \textit{error space} for a small number of extreme scenarios \cite{Sterpin_2021}. In OpenTPS, the dose is re-computed with MCsquare for 81 scenarios resulting from a combination of the setup and range scenarios previously described. The second strategy generates a confidence interval in the so-called \textit{dosimetric space}. It refers to all possible realizations of the treatment plan, or the different dose distributions that could be achieved. Described in \cite{Souris2019}, the scenarios are randomly sampled by MCsquare. The analysis in the dosimetric space reports the percentage of evaluation scenarios that meet a certain criterion such as minimum dose coverage for a target percentage (e.g., 95\% coverage for 90\% of scenarios). 

In OpenTPS, to proceed to robust evaluation, a Robustness object (associated to a plan) must be created and configured for the intented goal. A MCsquare simulation will compute the robust scenarios based on these information. The scenarios can be analyzed following the two strategies previously described using the member methods \textsf{analyzeErrorSpace} or \textsf{analyzeDosimetricSpace}. The DVHs of the nominal scenario and DVH bands can be recomputed for a list of ROIs and plotted for better visualization of the robustness of the generated treatment plan. 

\subsubsection{Robustness to intra-fraction motion (4D simulation)}
Two types of dose deterioration arise while treating moving targets in proton therapy using pencil beam scanning:
\begin{description}
    \item[Motion-induced range variations] The tumor motion can lead to a density variation in the beam path, thereby shifting the expected proton range and deteriorating the overall dose distribution.
    \item[Interplay effect] The tumor and scanning beam motions happen at the same time scale, leading to an interplay effect that can lead to hot and cold spots in the target.
\end{description}

To analyze those deteriorating effects, one can simulate the treatment delivery via Monte Carlo (MC) simulations on the 4DCT. In OpenTPS, those MC simulations are performed with MCsquare.

\paragraph{4D dose simulation}
To study the impact of the proton range variation, one can perform a 4D dose (4DD) simulation, which consists of independently simulating the dose delivery on each motion phase of the 4DCT and computing the average of those doses. Thus, each motion phase is considered static and contributes equally to the total dose. Mathematically the 4D dose is given by
\begin{align}
    \mathbf{F}_i &= \mathrm{DIR}(\mathbf{CT}_i,\mathbf{CT}_{\text{MidP}}) \text{ for all $i=1,...,P$} \\
    \mathbf{d}_{tot} &= \frac{1}{P}\sum_{i=1}^{P}\mathbf{d}^{\text{stat}}_i(\mathbf{v} + \mathbf{F}_i(\mathbf{v})) \text{ for all $\mathbf{v}\in\mathbb{R}^3$}
\end{align}
where $v$ is a voxel of the image, $P$ is the number of motion phases, $\mathbf{d}^{\text{stat}}_i$ is the dose resulting from the treatment plan simulation on the 3DCT of motion phase $i$, $\mathrm{DIR}(\cdot)$ is the deformable image registration operator, $\mathbf{F}_i$ is the deformation field resulting from the registration, $\mathbf{CT}_i$ is planning CT image $i$ and $\mathbf{CT}_{\text{MidP}}$ is the Mid-position CT. This resulting dose allows to analyze the impact of the proton range variation when motion is present and isolating it from the interplay effect.

\paragraph{4D dynamic dose simulation}
To study the impact of the interplay effect, one can perform a 4D dynamic dose (4DDD) simulation, which consists of simulating the delivery of beam spots one by one, via a time delivery model, on a sequence of 3DCTs. The sequence of 3DCTs is generally a looped version of the planning 4DCT, but it could be a richer set of synthetic CTs representing multiple breathing motions. The time delivery model is a model that associates a timing with the delivery of each spot in the treatment plan. This model is generally machine-specific and gives an expected delivery timing depending on several parameters such as the delay between the delivery of two adjacent spots, the time required for switching energy, etc. Knowing the delivery timing of each spot in the treatment plan, we can create a breathing model and know exactly in which motion phase each spot will be delivered. The breathing model has two parameters in the simplest case: the breathing period and the number of motion phases of the 4DCT. Mathematically the total dose is given by 
\begin{align}
    \mathbf{F}_i &= \mathrm{DIR}(\mathbf{CT}_i,\mathbf{CT}_{\text{MidP}}) \text{ for all $i=1,...,P$} \\
    \mathbf{d}_{tot} &= \sum_{i=1}^{P}\mathbf{d}^{\text{dyn}}_i(\mathbf{v} + \mathbf{F}_i(\mathbf{v})) \text{ for all $\mathbf{v}\in\mathbb{R}^3$.}
\end{align}
where $\mathbf{d}^{\text{dyn}}_i$ is the dose associated with the motion phase $i$ that includes all spots predicted to be delivered in that motion phase. Repeating the simulation by starting the delivery at different motion phases allows for quantifying the uncertainty of the final dose distribution. Note that we suppose each motion phase is a static image, thus neglecting motion during a motion phase.

\subsection{Data augmentation}

Moving targets, such as lung or liver tumors, come with an additional challenge: intra-fraction motion such as breathing, and usually higher inter-fraction motion than static target such as baseline shifts. To simulate treatment delivery on this kind of tumors, tools have been added to OpenTPS to add intra- and inter-fraction motions, as well as creating training and test data based on motion models (see section \ref{dataManagAndProc}) to evaluate innovative methods or AI models. 
\\
\paragraph{Inter-fraction motion simulation}

Four types of inter-fraction motion can be artificially generated.

\begin{description}
    \item[Baseline shifts]
Artificial baseline shifts can be applied on \textsf{3DCT} or \textsf{Dynamic3DModels} based on the \textsf{ROIMask} of the target or organ to be translated. The baseline shift is constructed in the form of a diffeomorphic displacement vector field modeling a local shift of the \textsf{ROIMask} while preserving surrounding bony structures.
    \item[Organ/target shrinks]
Organ or target shrinks can be applied on \textsf{Image3D} or \textsf{Dynamic3DModel} objects, using a \textsf{ROIMask} for the organ that must be shrinked. First, the \textsf{ROIMask} is dilated by 1 voxel and eroded by a number of pixels depending on the user input in mm. From these two new masks, the dilated and eroded bands corresponding to the difference with the original mask are computed. Then, one by one, every voxel of the eroded band is replaced by a random sample picked from a normal distribution $N(\mu, \sigma ^{2})$ where $\mu$ is the average value of the 10 voxels from the dilated band closest to the voxel to fill.
    \item[Rotations]
Rotations around the 3 main axis can be applied to \textsf{Image3D} or \textsf{Dynamic3DModel} objects. The rotation is applied around an axis at the center of the image and not around the image origin. Note that in case of multiple rotations, the order is important as rotations are not commutative.
    \item[Translations]
Translations in the 3 main axis can be applied to \textsf{Image3D} or \textsf{Dynamic3DModel} objects. 
\end{description}

\begin{figure*} 
\centering
\includegraphics[width=\textwidth]{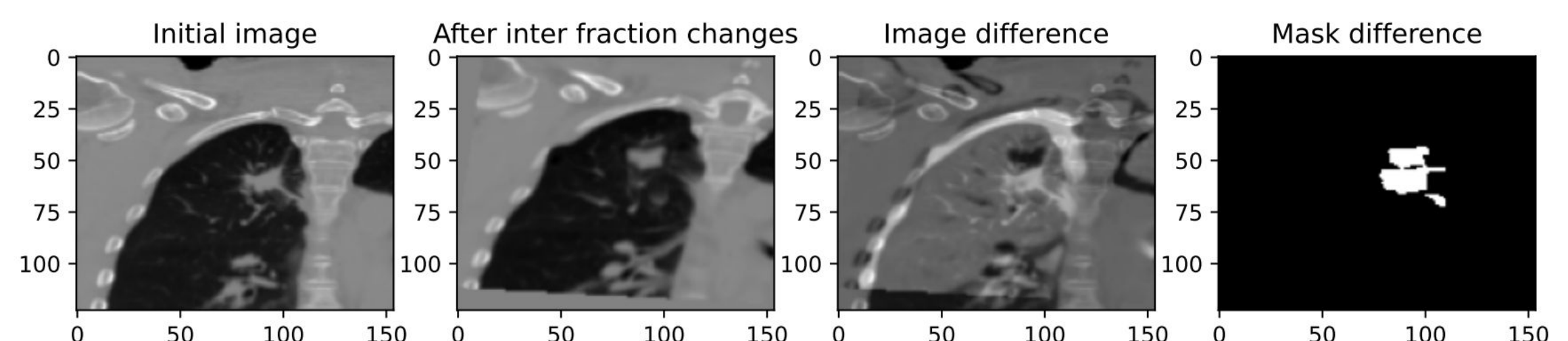}
\caption{Example of inter-fractions transforms applied to a 3DCT. Small translations and rotations were applied to the entire image, and additionally a baseline shift and a shrink were applied to the lung tumor.}
\label{interFrac}
\end{figure*}

An example of synthetic inter-fraction modification on a \textsf{3DCT} image can be seen in Figure \ref{interFrac}.

\paragraph{Intra-fraction breathing simulation}

\begin{figure} 
\centering
\includegraphics[width=0.85\textwidth]{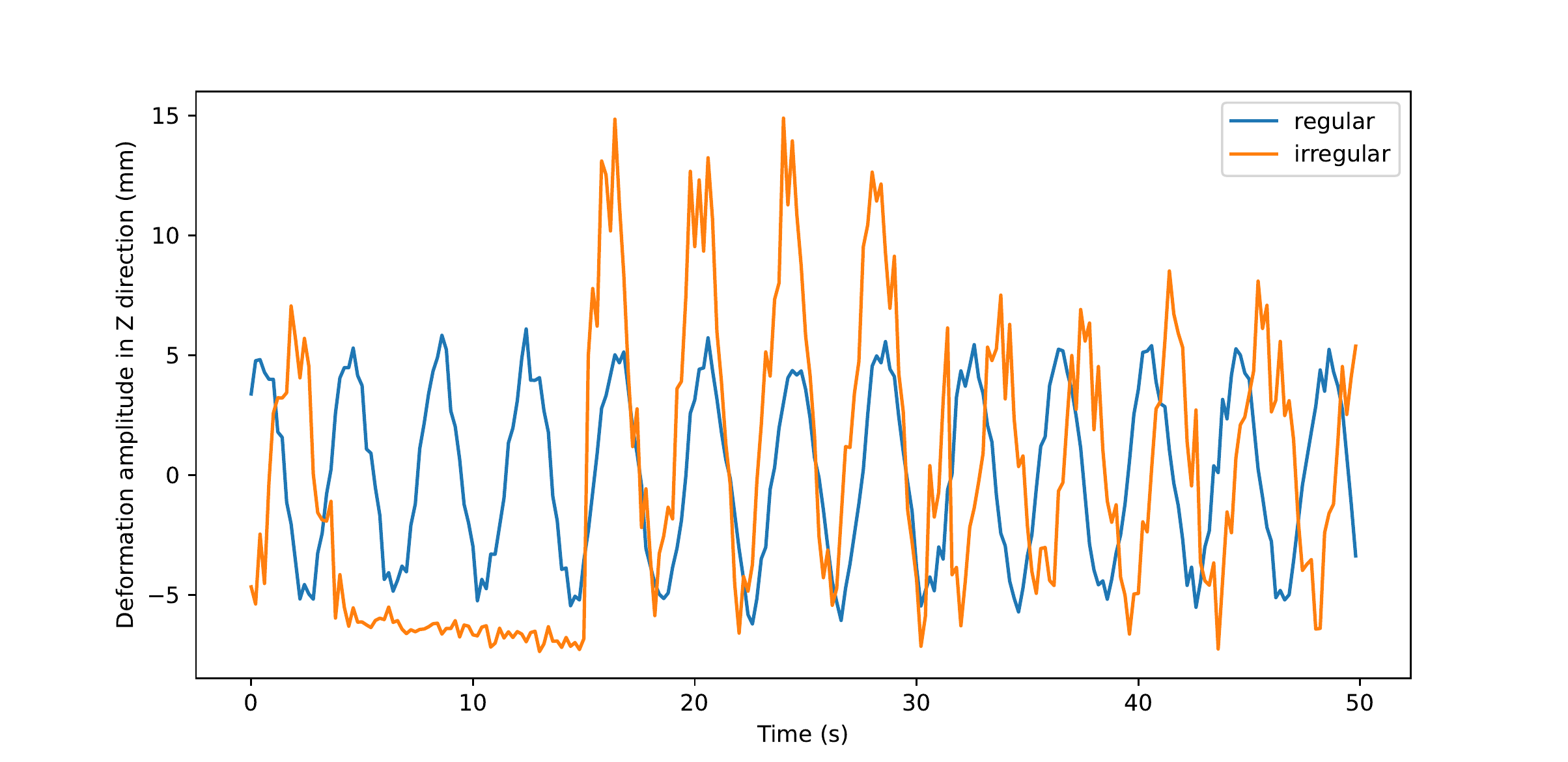}
\caption{Example of synthetic breathing signals, regular in blue and irregular in orange.}
\label{breathingSig}
\end{figure}

Using a motion model, a sequence of images following a specific breathing pattern can be created. First, synthetic 1D or 3D breathing signals can be generated with different parameters to change the breathing period, the noise or to add irregularities such as apneas (Figure \ref{breathingSig}).

Then, one or multiple points can be selected in the midP image of a motion model to apply these breathing signals in a specified direction. If multiple points and signals are used (for example to partially decorrelate the motion of the tumor and the motion of the skin), they are combined using weight maps and linear interpolations \cite{DasnoySumell2022} to create 3D image sequences containing any type of breathing patterns. These image sequences can be used to evaluate treatment robustness, new treatment methods \cite{hamaide2022real} or create additional patient data to train AI methods \cite{loyen20213dct}.

\section{Perspectives}

OpenTPS is a unique platform for research in proton therapy treatment planning, being an open-source tool written in a user-friendly programming language. This allows for both internal contributions and international collaboration to enhance the platform quickly, making it increasingly sophisticated over time. Our research group is currently focusing on several exciting areas within OpenTPS, including arc proton therapy treatment planning \cite{WUYCKENS2022105609,Wuyckens_2022} and FLASH proton therapy\cite{Deffet2023}. Depending on the needs of the users, new features might be added in the future such as (but not limited to) other treatment or image modalities, objects tracking or position prediction in dynamic sequences. Additionally, the platform has potential for real-time treatment delivery visualization and 2D to 3D image reconstruction. As the field of proton therapy continues to evolve, OpenTPS offers a flexible and dynamic platform for exploring new possibilities and pushing the boundaries of what is possible.

The information to get started are available on the website\footnote{\url{http://opentps.org/}} or on gitlab\footnote{\url{https://gitlab.com/openmcsquare/opentps}}.


\section{Acknowledgments}
During the development of OpenTPS, Dr. Sylvain Deffet received funding of the Walloon Region of Belgium through technology innovation partnership no. 8341 (EPT-1 – Emerging Proton Theraexact Phase 1) co-led by MecaTech and BioWin clusters. Sophie Wuyckens and Kevin Souris are funded by the Walloon Region as part of the Arc Proton Therapy convention (Pôles Mecatech et Biowin). Dasnoy Damien, Gauthier Rotsart de Hertaing and Valentin Hamaide are supported by the Walloon Region, SPWEER Win2Wal program project 2010149. Estelle Loÿen is a Televie grantee of the Fonds de la Recherche Scientifique - F.N.R.S. Computational resources have been provided by the supercomputing facilities of the Université catholique de Louvain (CISM/UCL) and the Consortium des Équipements de Calcul Intensif en Fédération Wallonie Bruxelles (CÉCI) funded by the F.R.S.-FNRS under convention 2.5020.11.

\bibliographystyle{medphy}
\bibliography{biblio}

\end{document}